\documentclass[twocolumn]{emulateapj}
\usepackage{apjfonts}

\newcommand{\eg}{\emph{e.g.}}

\newcommand{\be}{\begin{equation}}
\newcommand{\ee}{\end{equation}}
\newcommand{\hc}{HC$_3$N}
\newcommand{\hct}{H$^{13}$CCCN}
\newcommand{\hcf}{HCCC$^{15}$N}
\newcommand{\ratio}{HC$_3$N/HCCC$^{15}$N}
\newcommand{\ration}{$^{14}$N/$^{15}$N}

\submitted{ApJL, accepted for publication May 2018}

\shortauthors{Cordiner et al.}

\usepackage{graphicx}	% Including figure files
\usepackage{amsmath}	% Advanced maths commands
\usepackage{amssymb}	% Extra maths symbols

\begin{document}

\title{Interferometric imaging of Titan's HC$_3$N, H$^{13}$CCCN and HCCC$^{15}$N}

\author{M. A. Cordiner\altaffilmark{1,2}, C. A. Nixon\altaffilmark{1}, S. B. Charnley\altaffilmark{1}, N. A. Teanby\altaffilmark{3}, E. M. Molter\altaffilmark{4}, Z. Kisiel\altaffilmark{5}, V. Vuitton\altaffilmark{6}}

\altaffiltext{1}{NASA Goddard Space Flight Center, 8800 Greenbelt Road, MD 20771, USA.}
\email{martin.cordiner@nasa.gov}
\altaffiltext{2}{Institute for Astrophysics and Computational Sciences, The Catholic University of America, Washington, DC 20064, USA.}\
\altaffiltext{3}{School of Earth Sciences, University of Bristol, Wills Memorial Building, Queen's Road, Bristol, BS8 1RJ, UK.}
\altaffiltext{4}{Astronomy Department, 501 Campbell Hall, University of California Berkeley, Berkeley, CA, USA.}
\altaffiltext{5}{Institute of Physics, Polish Academy of Sciences, Al. Lotnik{\'o}w 32/46, 02-668 Warszawa, Poland.}
\altaffiltext{6}{Universit{\'e} Grenoble Alpes, CNRS, IPAG, F-38000 Grenoble, France.}

\begin{abstract}
We present the first maps of cyanoacetylene isotopologues in Titan's atmosphere, including H$^{13}$CCCN and HCCC$^{15}$N, detected in the 0.9~mm band using the Atacama Large Millimeter/submillimeter array (ALMA) around the time of Titan's (southern winter) solstice in May 2017. The first high-resolution map of HC$_3$N in its $v_7=1$ vibrationally excited state is also presented, revealing a unique snapshot of the global HC$_3$N distribution, free from the strong optical depth effects that adversely impact the ground-state ($v=0$) map.  The HC$_3$N emission is found to be strongly enhanced over Titan's south pole (by a factor of 5.7 compared to the north pole), consistent with rapid photochemical loss of HC$_3$N from the summer hemisphere combined with production and transport to the winter pole since the April 2015 ALMA observations.  The H$^{13}$CCCN/HCCC$^{15}$N flux ratio is derived at the southern HC$_3$N peak, and implies an HC$_3$N/HCCC$^{15}$N ratio of $67\pm14$. This represents a significant enrichment in $^{15}$N compared with Titan's main molecular nitrogen reservoir, which has a \ration\ ratio of 167, and confirms the importance of photochemistry in determining the nitrogen isotopic ratio in Titan's organic inventory.

\end{abstract}


%%%%%%%%%%%%%%%%%%%%%%%%%%%%%%%%%%%%%%%%%%%%%%%%%%

%%%%%%%%%%%%%%%%% BODY OF PAPER %%%%%%%%%%%%%%%%%%

\section{Introduction}
Titan's nitrogen and methane-dominated atmosphere is by-far the densest of any moon in the Solar System, and its origin has remained a mystery since its discovery by \citet{kui44} --- see \citet{hor17} and references therein. Titan's atmospheric nitrogen is theorized to have been outgassed as NH$_3$ or N$_2$, originally present as ice that was accreted in the Saturnian sub-nebula, perhaps with a contribution delivered by cometary impacts. Such theories can be tested by measurements of Titan's present-day atmospheric abundances.

Trace isotopic ratios can reveal crucial information on the origins of a wide variety of solar system materials \citep[\eg][]{man09,boc15,mar18,ale18}, and provide unique insights into their thermal and chemical histories. The gases that were eventually incorporated into icy planetesimals may have become enriched (or depleted) in heavy isotopes during the formation of the solar system or prior interstellar cloud. The difference in zero point energy between the reactants and products means that at low temperatures (below the activation energy for the reverse reaction), isotopic exchange reactions such as $^{15}$N$^+$ + $^{14}$N$_2$ $\leftrightharpoons$ $^{14}$N$^+$ + $^{14}$N$^{15}$N tend to procede preferentially in the forward direction \citep[see][]{rou15}. \citet{cha02} theorized that reactions of $^{15}$N-enriched N$_2$ with He$^+$, followed by H$_2$ can lead to the production of $^{15}$N-enriched interstellar NH$_3$ ice. In the protosolar nebula, $^{15}$N-enriched NH$_3$ may have also arisen following the isotope-selective photodissociation of N$_2$ \citep[see][]{vis18}. Understanding Titan's atmospheric \ration\ ratio may thus provide a crucial window into the thermal, chemical and radiation history of its nitrogen-bearing ices. 

The first measurement of Titan's \ration\ ratio was by \citet{mar02}, who used mm-wave spectroscopy of HCN to derive HC$^{14}$N/HC$^{15}$N = 60-70. A small ratio (compared to the Solar value of 440 and the terrestrial value of 272) was confirmed by \citet{gur04}, \citet{vin07b} and \citet{cou11}, using a combination of ground and space-based sub-mm and infrared observations. A refined disk-average measurement of $72.3\pm2.2$ was recently obtained by \citet{mol16} using ALMA archival flux-calibration observations of Titan. Meanwhile, the Cassini-Huygens mass spectrometer measured the \ration\ ratio in tropospheric N$_2$ to be significantly higher at $167\pm0.6$ \citep{nie10}. The difference in $^{15}$N fraction for these molecules is theorized to be a consequence of isotope-selective photodissociation of N$_2$ in the upper atmosphere. At altitudes $\gtrsim800$~km, $^{14}$N$_2$ is more slowly dissociated than $^{15}$N$^{14}$N due to self-shielding in the predissociating absorption lines \citep{lia07}. This gives rise to an enhanced abundance of atomic $^{15}$N that is theorized to carry through into other photochemically-produced species. While this is likely sufficient to explain the observed HC$^{14}$N/HC$^{15}$N ratio, the theory remains to be tested for any molecules apart from HCN. 

Additional complexity arises due to the many possible sources and sinks of atmospheric $^{14}$N and $^{15}$N, including outgassing from (or precipitation onto) the surface, and sputtering/escape from (delivery to) the top of the atmosphere, any of which may alter the overall nitrogen isotopic ratio over time \citep{man09,kra16}. Our present study is motivated by the need to accurately measure the \ration\ ratio in Titan's photochemical products, to help elucidate the sources and sinks of $^{15}$N, which are crucial for a proper understanding of the primordial value of \ration\ in Titan's ice at the time it was accreted. Stratospheric \hc\ densities are theorised to be affected by many different reactions, so comparison between observed and predicted HCCC$^{15}$N abundances provides a crucial check of the reaction networks used in photochemical models \citep[\eg][]{loi15,kra16,vui18}, including photolysis cross-sections, reaction rates and branching ratios.

In contrast to $^{15}$N, a large body of remote and in-situ observational (and laboratory) data shows that $^{13}$C ratios are much less variable, with a value of $\approx90$ across a wide range of solar system materials (including all of Titan's hydrocarbons and nitriles for which measurements exist; \citealt{bez14}). This implies either that no significant carbon fractionating processes are in operation, or that the isotopic production and loss mechanisms are closely balanced \citep{hor17}. 

In this study, we present the first maps of Titan's $^{13}$C and $^{15}$N isotopologues of the cyanoacetylene molecule (\hc), obtained using spatially resolved data from the Atacama Large Millimeter/submillimeter Array (ALMA). Pure mm/sub-mm rotational emission lines of the main H$^{12}$C$_3$$^{14}$N isotopologue (hereafter referred to as simply \hc), are found to be unreliable as a tracer of the total \hc\ abundance due to strong opacity effects, so we use the known atmospheric $^{12}$C/$^{13}$C ratio (assumed to apply to \hc), combined with our ALMA measurements of \hct\ and \hcf\ to derive the \ration\ ratio in \hc\ for the first time.

\section{Observations}

\begin{table*}
\centering
\caption{Detected HC$_3$N line spectroscopic parameters and measured fluxes \label{tab:results}}
\begin{tabular}{llccccl}
\hline\hline
Species &    Transition     &Rest Freq.  &$A$   &$g_u$& $E_u$ & Flux\\
        &                   &(MHz)       &($10^{-3}$\,s)&&  (K)  & (Jy\,kHz)\\
\hline
HC$_3$N & $J=39-38,\,v=0$    & 354697.463 & 3.571 & 79 & 340.5 & $475\pm19$\\
HC$_3$N & $J=39-38,\,v_7=1e$ & 355566.254 & 3.577 & 79 & 662.2 & $412\pm17$\\
HC$_3$N & $J=39-38,\,v_7=1f$ & 356072.445 & 3.592 & 79 & 662.7 & $402\pm17$\\
H$^{13}$CCCN & $J=39-38$      & 343737.400 & 3.250 & 79 & 330.0 & $35.5\pm5.9$\\
HCCC$^{15}$N & $J=39-38$      & 344385.348 & 3.254 & 79 & 330.6 & $47.3\pm6.6$\\
\hline
\vspace*{2mm}
\end{tabular}
\end{table*}

Observations of Titan were obtained using ALMA on 2017-05-08 as part of the Director's Discretionary Time program 2016.A.00014.S. Following the initial bandpass and flux calibration scans, our observations consisted of an interleaved sequence of three visits each to Titan and the phase calibrator J1751-1950. The phase-center was updated in real-time to track Titan's moving position on the sky.

The Band 7 receiver was used, and the ALMA correlator was configured to observe the frequency ranges 342.5-346.1~GHz and 354.2-356.1~GHz at moderate spectral resolution (976~kHz) to capture the CO 2-1 and HCN 4-3 lines (including their broad line wings), as well as the HC$_3$N lines of interest to this study. The total on-source observing time for Titan was 18 min, with 46 antennas online, resulting in an RMS noise of $\approx4$~mJy\,beam$^{-1}$. Weather conditions were good, with a zenith precipitable water vapor of 0.76 mm.

Data were flagged and calibrated in CASA 5.1 \citep{jae08} using the automated pipeline scripts supplied by the Joint ALMA Observatory \citep[see][]{shi15}. Flux calibration was performed with respect to the quasar J1733-1304, and is expected to be accurate to within about 5\%. The spectral axis was transformed to Titan's rest frame and regridded to a 976~kHz channel width. Titan's continuum flux was subtracted using low-order polynomial fits to the spectral regions adjacent to our detected lines. Imaging and deconvolution were performed using the {\tt Clark clean} algorithm with natural visibility weighting, a pixel size of $0.025''$ and a flux threshold of 8 mJy. The resulting angular resolution was $0.23''\times0.17''$ from a Gaussian fit to the point-spread function.

The coordinate scales of the cleaned images were transformed to physical distances with respect to the center of Titan, which was 9.26~au from Earth at the time of observation. Titan's north pole was oriented $5.3^{\circ}$ counter-clockwise from celestial north, and tilted towards the observer by $26^{\circ}$. This is close to the maximum polar tilt due to the proximity of our observations to Titan's southern winter solstice on 2017 May 24.  Before plotting, our images were corrected for a small, $0.06''$ offset in declination (of unknown origin) that was identified in Titan's position with respect to the ALMA phase center.

\section{Results}

Emission from HC$_3$N, including lines from the ground ($v=0$) vibrational state, ($v_7=1$) vibrationally excited state, and the isotopologue lines were identified using the CDMS catalogue \citep{mul01}, based on the laboratory frequencies of \citet{tho00,tho01}. Relevant spectroscopic parameters and integrated line fluxes are given in Table \ref{tab:results}. Note that although the $J=38-37$ lines of HC$^{13}$CCN and HCC$^{13}$CN were detectable in addition to the $J=39-38$ line of \hct, the other $^{13}$C isotopologues were excluded from the present study as they cannot be properly disentangled from the steeply-rising wing of the overlapping HC$^{15}$N $J=4-3$ line (at 344200~MHz).

\begin{figure*}
\centering
\includegraphics[width=0.4\textwidth]{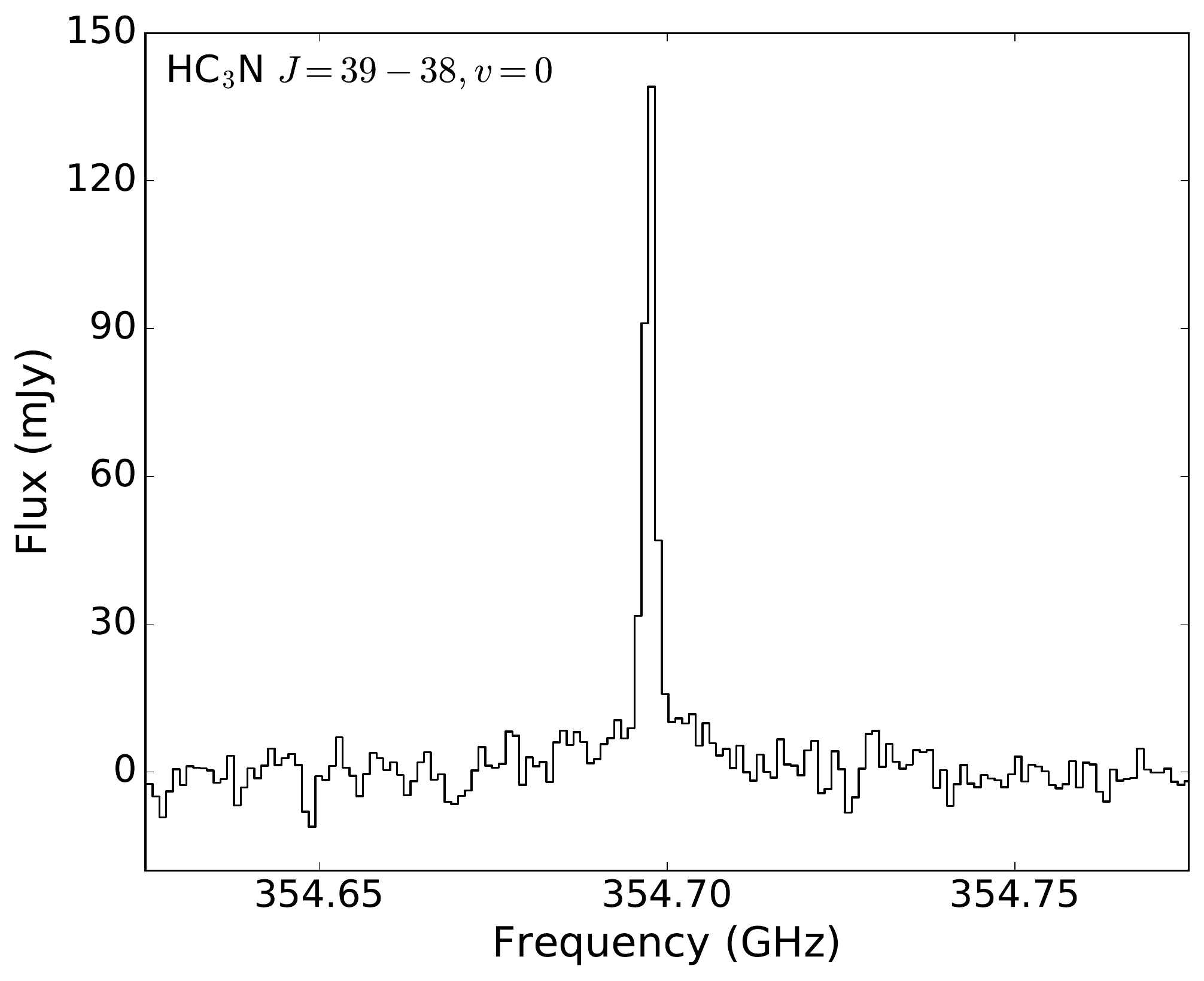}
\hspace{3mm}
\includegraphics[width=0.4\textwidth]{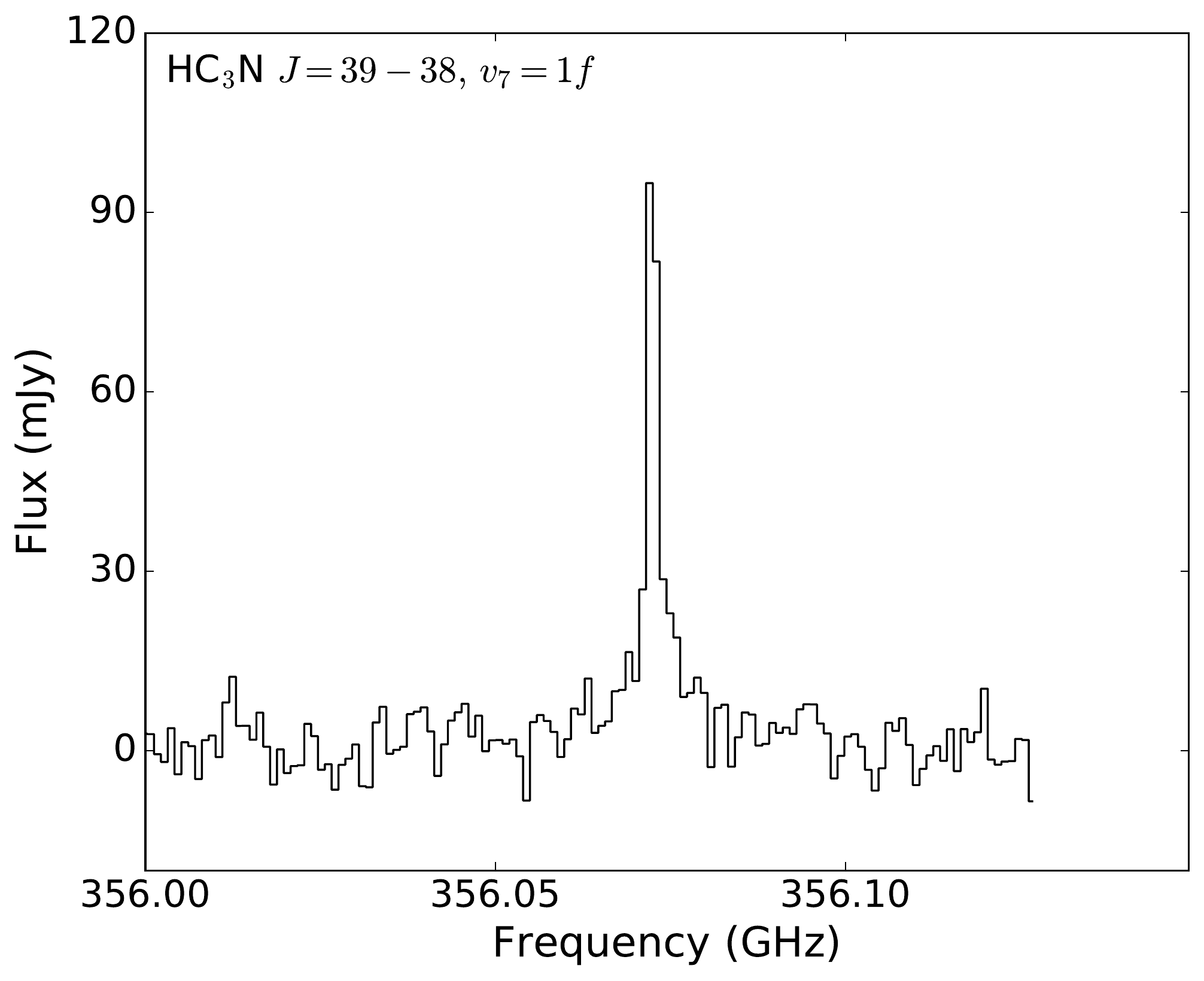}\\
\includegraphics[width=0.4\textwidth]{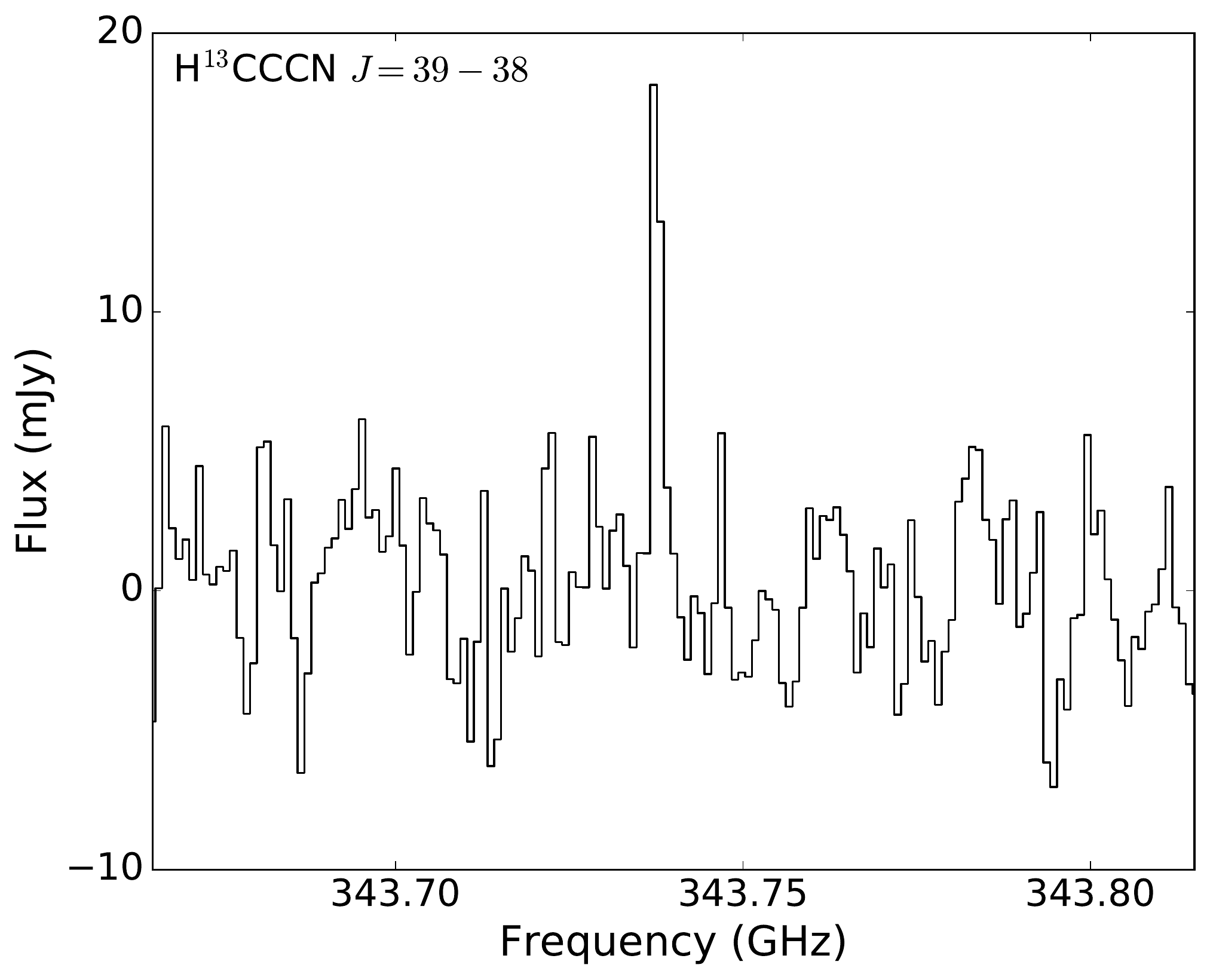}
\hspace{3mm}
\includegraphics[width=0.4\textwidth]{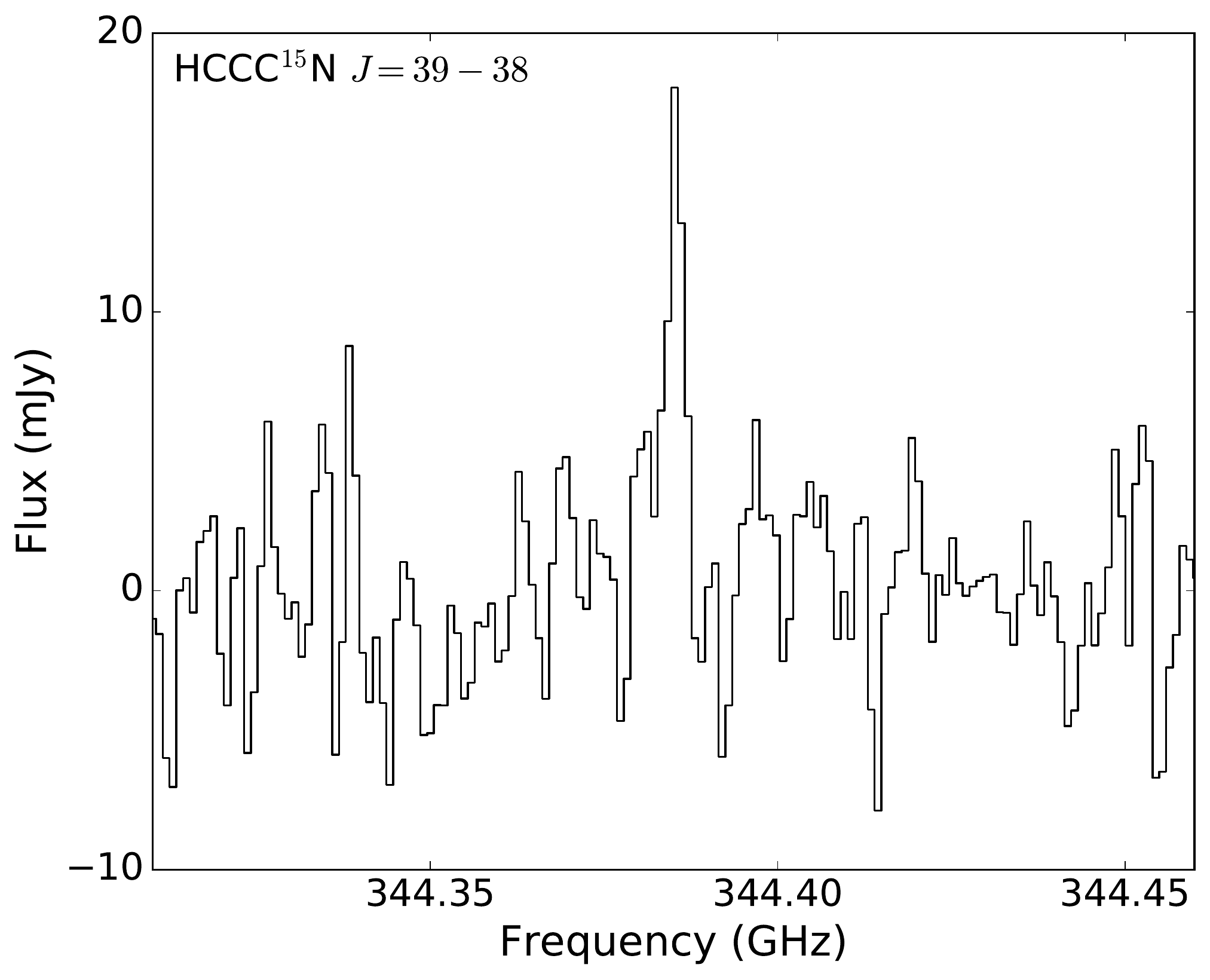}
\caption{ALMA spectra of HC$_3$N extracted from a beam centered 200~km above the south polar limb, including the ($J=39-38,\,v=0$) rotational transition of the vibrational ground state, the ($J=39-38,\,v_7=1f$) transition of the vibrationally excited state, and the ground-state rotational transitions of the detected $^{13}$C and $^{15}$N isotopologues (see Table \ref{tab:results}). \label{fig:spec}}
\end{figure*}

\begin{figure*}
\includegraphics[width=0.49\textwidth]{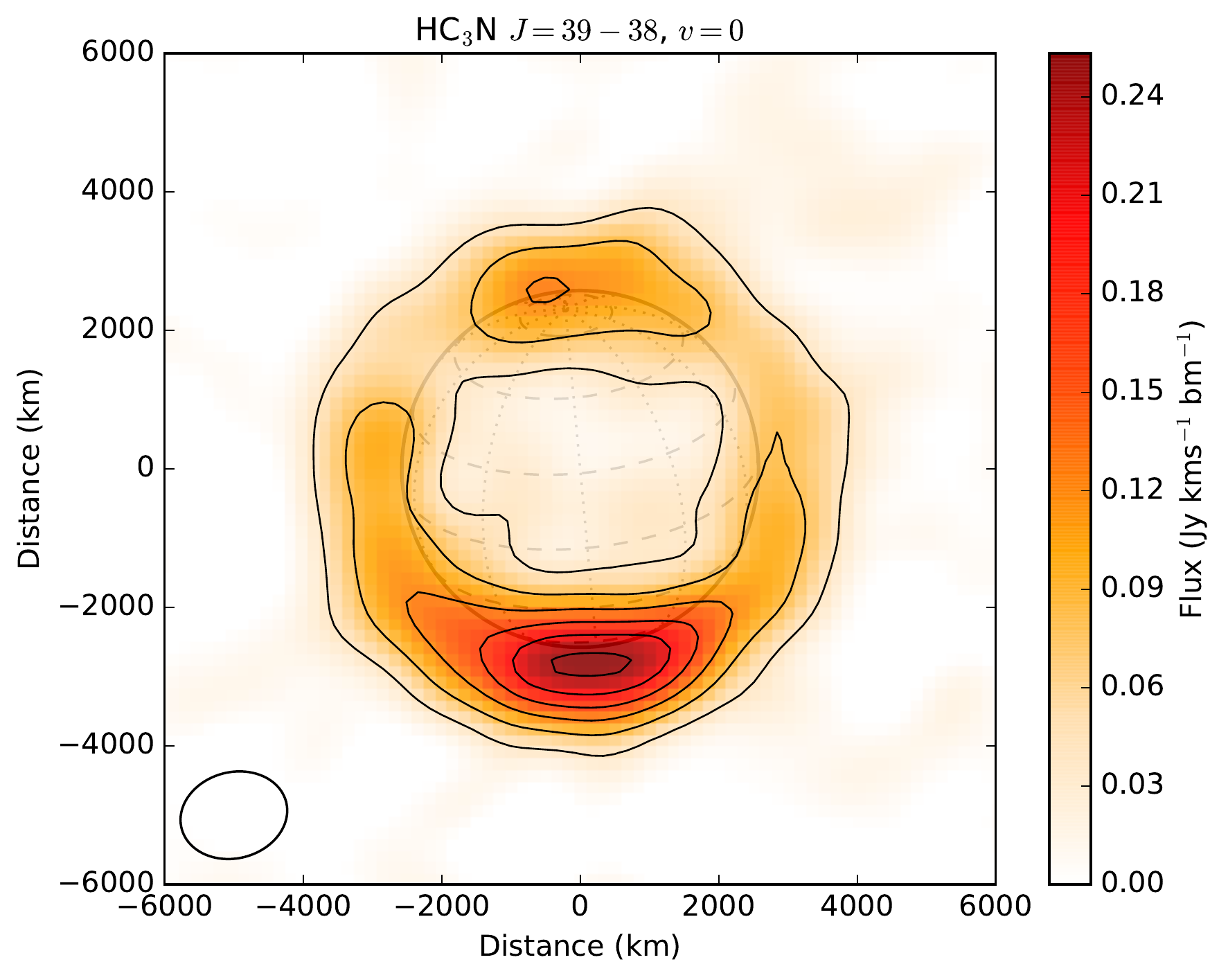}
\includegraphics[width=0.49\textwidth]{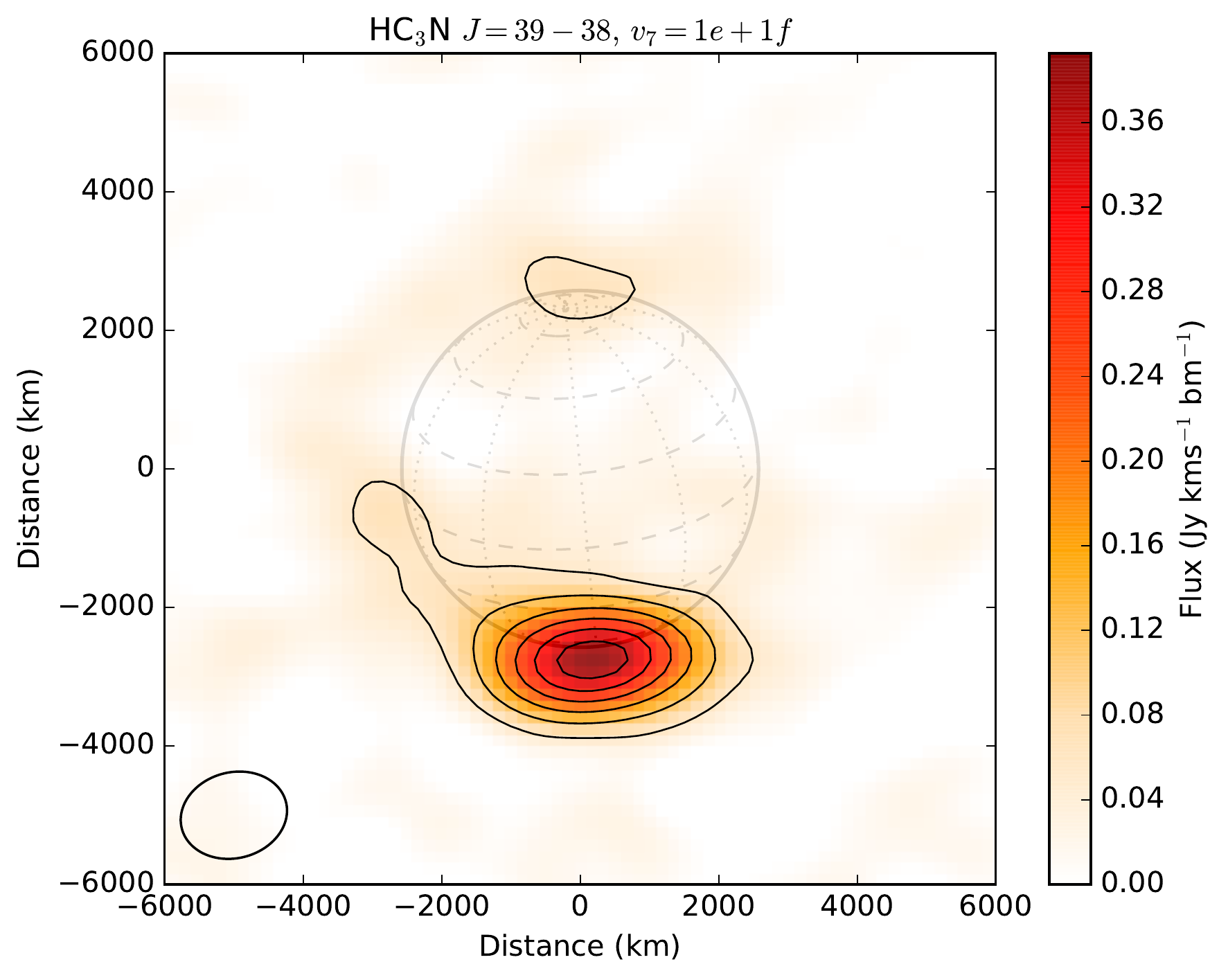}\\
\includegraphics[width=0.49\textwidth]{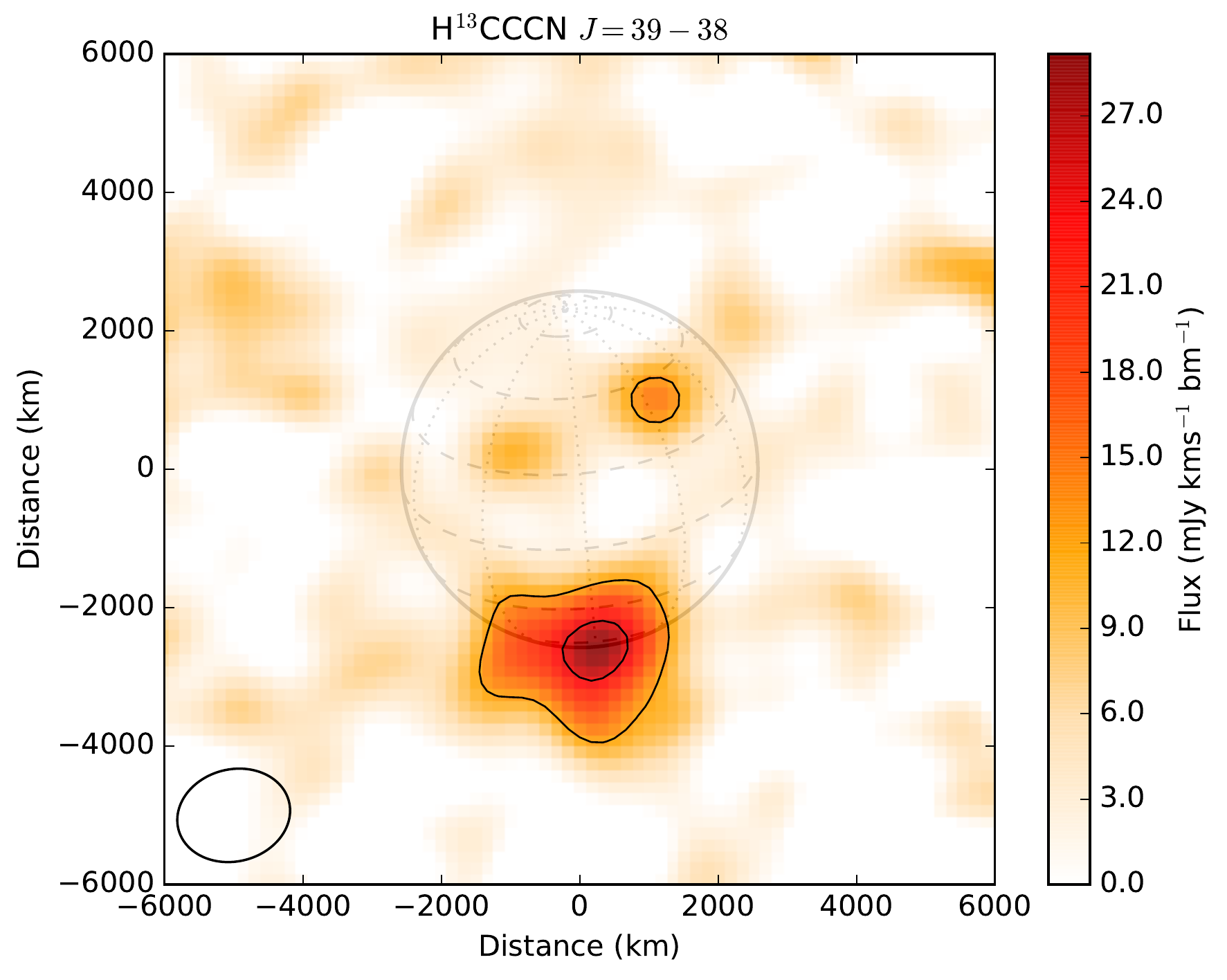}
\includegraphics[width=0.49\textwidth]{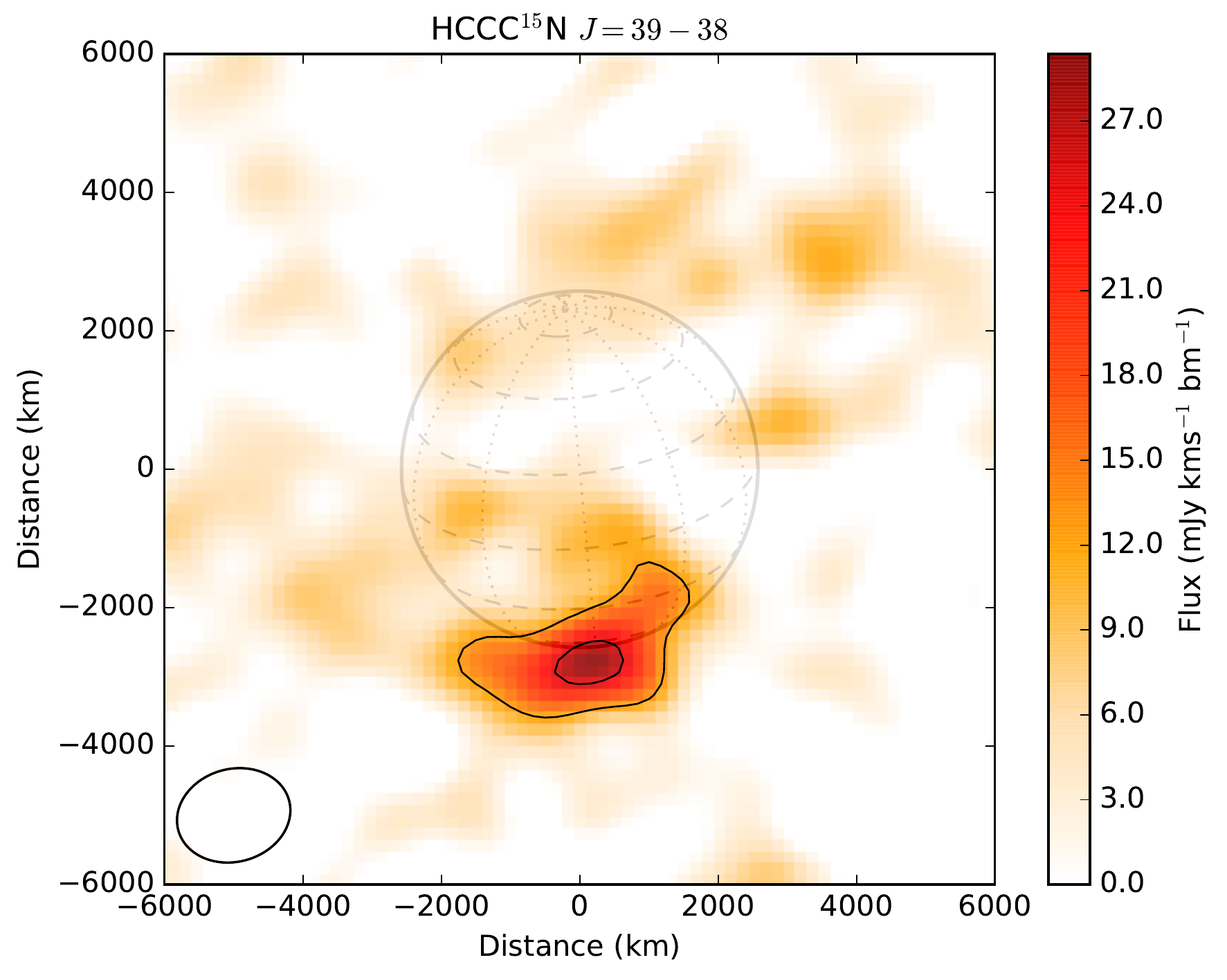}
\caption{Integrated emission maps of HC$_3$N, including the ground-state ($J=39-38,\,v=0$) and vibrationally excited ($J=39-38,\,v_7=1e$ and $1f$) lines (top two panels), and the detected $^{13}$C and $^{15}$N isotopologues (bottom two panels). The contour interval is  $5\sigma$ for the HC$_3$N $v=0$ and $v_7=1$ lines, and $3\sigma$ for the isotopologues, where $\sigma$ is the RMS noise level. Wire frame shows Titan's solid body, with $22.5^{\circ}$ increments in latitude and $30^{\circ}$ in longitude. Ellipses (lower left) indicate the spatial resolution. \label{fig:maps}}
\end{figure*}

Figure \ref{fig:spec} shows the detected spectral lines and Figure \ref{fig:maps} shows the emission maps, integrated over the full extent of the detected flux for \hct\ and \hcf. Only the central 5 spectral channels were included for HC$_3$N $v=0$ and $v_7=1$ to facilitate intercomparison of their maps --- avoiding the introduction of undue noise from the weak, pressure-broadened wings that are present at the north pole for $v=0$ (but not detected for $v_7=1$). \hc\ ($v=0$) shows a limb-brightened flux distribution, characteristic of high-altitude atmospheric emission. There is a strong intensity peak over the south pole, with a weaker, secondary peak in the north. By contrast, emission from the $^{13}$C and $^{15}$N isotopologues was only detected within a compact region over the south pole, with a peak flux $0.03''$ away from Titan's disk. This corresponds to a sky-projected altitude of 200~km above the southern limb.

The Einstein A coefficients and degeneracies ($g_u$) for the $v=0$ and $v_7=1$ lines are practically identical, but the upper-state energy ($E_u$) for $v_7=1$ is 322~K above $v=0$. At the $\sim160$~K temperature in the middle atmosphere where the majority of the \hc\ emission originates, the difference in Boltzmann factors results in a factor of 7.4 reduction in the population of the vibrationally-excited state (assuming local thermodynamic equilibrium; LTE), with a corresponding drop in emitted line flux. In fact, the measured $v=0$ line flux is only 1.2 times stronger than $v_7=1$, and this discrepancy is probably due to a very high optical depth of the $v=0$ line in a compact region near the south pole (see section \ref{sec:dis}). Expressed as a Rayleigh-Jeans antenna temperature, the HC$_3$N $v=0$ line peak of 78 K does not expose the full extent of the line saturation, which is more readily revealed by comparing the emission from the north (N) and south (S) poles. The ratio of S to N polar peaks for $v_7=1$ is $5.7\pm0.3$, and for $v=0$ the ratio is only $1.6\pm0.1$. This difference cannot be attributed to differences in temperatures between the poles, which are typically $\lesssim20$~K --- such a temperature variation only amounts to $\sim20\%$ difference in the LTE $v_7=1$ and $v=0$ level populations. Thus, we deduce the presence of a south polar region smaller than the telescope beam, in which the HC$_3$N  $J=39-38,\,v=0$ line is completely optically thick, significantly depressing the measured flux from that region.

The high opacity of the HC$_3$N  $J=39-38,\,v=0$ line rules out its use for deriving an accurate abundance over the south pole where the isotopologues were detected. We have also conducted tests to retrieve the \hc\ abundance using the $v_7=1$ transition, but this line is found to be too sensitive to errors in the adopted temperature profile. Moreover, it has recently been shown that, due to the rapidly decreasing density, non-LTE effects start to become important for vibrationally excited lines in Titan's atmosphere above about 350~km \citep{kut13}, resulting in unreliable abundance retrievals. Given that a significant proportion of the \hc\ flux detected through high-resolution mm/sub-mm spectroscopy originates from altitudes above 350~km \citep{mar02,cor14}, we find ourselves in the unfortunate situation of being unable to derive accurate enough abundances for the main H$^{12}$C$_3$$^{14}$N isotopologue to allow a direct calculation of the $^{12}$C/$^{13}$C and \ration\ ratios in this molecule.

A similar situation is encountered for HCN in the interstellar medium, and can plausibly be resolved using the `double isotope' method: the more easily measured abundance of an (optically thin) $^{13}$C-substituted isotopologue is combined with the known $^{12}$C/$^{13}$C ratio to infer the abundance of the (optically thick) $^{12}$C isotopologue. The use of this method for Titan's \hc\ rests on the assumption that the \hc/\hct\ ratio is the same as the bulk $^{12}$C/$^{13}$C ratio measured from other gases.

The $^{12}$C/$^{13}$C ratios across the planets, moons and minor bodies of the solar system tend to cluster around 90 \citep{wood09}.  Similarity between the $^{12}$C/$^{13}$C ratios for Jupiter, Saturn, the Earth and Sun led \citet{sad96} to conclude that there is little or no $^{13}$C fractionation occurring in the atmospheres of the Giant Planets. The range of individual $^{12}$C/$^{13}$C measurements for Titan's gases tabulated by \citealt{bez14} are all consistent with the (error-weighted) average value of $88.6\pm0.8$. This includes the prior measurement of $79\pm17$ in \hc\ by \citet{jen08}, and we take this as good evidence for a lack of strong $^{13}$C fractionation processes operating in Titan's atmosphere.

Since the molecular partition functions, Einstein A values and upper-state energies for the observed \hct\ and \hcf\ transitions are identical to within 0.2\%, the \hct/\hcf\ abundance ratio can be accurately derived, independent of assumptions regarding the temperature and excitation of the gas. Adopting a value H$^{12}$C$_3$$^{14}$N/H$^{13}$C$^{12}$C$_2$$^{14}$N = 89, the ratio of our observed \hct\ and \hcf\ line fluxes ($0.75\pm0.16$) implies a \ration\ ratio of $67\pm14$ in cyanoacetylene. The measured fluxes are dominated by emission from the (narrow) spectral line cores, which originates primarily from altitudes $\sim200-400$~km \citep[\eg][]{mar02}, so the measured \ration\ ratio should be considered an average over this range.\\[3mm]

\section{Discussion}
\label{sec:dis}

Our value for \ration\ in \hc\ is consistent with the average of the prior HCN measurements: $69.3\pm3.4$ \citep{bez14,mol16}. This may be understood within the framework of recent models for Titan's nitrile photochemistry. A source of $^{15}$N-enriched atomic nitrogen is produced in the upper atmosphere as a result of preferential photodissociation of $^{15}$N$^{14}$N compared with the dominant $^{14}$N$_2$ isotopologue. The line radiation required to predissociate $^{14}$N$_2$ becomes attenuated with distance into the atmosphere, and this effect is weaker for $^{15}$N$^{14}$N due to its lower abundance and differing predissociation wavelengths compared with the main isotopologue \citep{lia07}. The resulting $^{15}$N-enriched atomic nitrogen becomes incorporated into HCN, and then into \hc\ (through the CN radical intermediary), as highlighted by the following reaction sequence \citep[cf.][]{wil04}:

\be
{\rm ^{15}N^{14}N} + h\nu \longrightarrow {\rm ^{15}N} + {\rm ^{14}N} 
\ee

\be
{\rm ^{15}N} + {\rm CH_3} \longrightarrow {\rm H_2C^{15}N} + {\rm H} 
\ee

\be
{\rm H_2C^{15}N} + {\rm H} \longrightarrow {\rm HC^{15}N} + {\rm H_2} 
\ee

\be
{\rm HC^{15}N} + h\nu \longrightarrow {\rm C^{15}N} + {\rm H} 
\ee

\be
{\rm C^{15}N} + {\rm C_2H_2} \longrightarrow {\rm {HC_3}^{15}N} + {\rm H} 
\ee

Given the expected altitude dependence of the atomic \ration\ ratio (as $^{15}$N$^{14}$N is photodissociated to greater depths than $^{14}$N$_2$), and the possibility of multiple formation and destruction pathways for \hc\ \citep[\eg][]{loi15}, which become important in different parts of the atmosphere, the \ration\ ratio in \hc\ cannot be easily interpreted without the aid of a detailed chemical model. The latest models by \citet{vui18} and \citet{dob18} incorporate a comprehensive nitrile chemistry, including $^{15}$N-bearing species as well as ion, cosmic ray and photolytic processes. \citet{vui18} are able to reproduce the previously-obtained HCN/HC$^{15}$N ratio from Cassini and predict HC$_3$N/HCCC$^{15}$N = 52 at 200 km, which is in reasonably good agreement with our ALMA observations. \citet{dob18} predicted a higher \ratio\ ratio of $80\pm7$ by including dissociation by energetic electrons from Saturn's magnetosphere.  Unfortunately, our measured \ratio\ ratio is not accurate enough to conclusively distinguish between these models, so additional observations at higher sensitivity are needed. Another source of uncertainty stems from our assumption of H$^{12}$C$_3$N/H$^{13}$CCCN = 89; limited accuracy of the prior $^{12}$C/$^{13}$C measurement in \hc\ \citep{jen08} means that the true \ratio\ ratio could be as low as 43, so more accurate measurements of HC$_3$N/H$^{13}$CCCN are also warranted. Combined with information on the $^{12}$C/$^{13}$C ratios in C$_2$H$_2$, HCN and other molecules, ALMA observations of the remaining $^{13}$C isotopologues HC$^{13}$CCN and HCC$^{13}$CN would allow the individual $^{13}$C atoms to be tracked through the chemical network, providing a unique test for our understanding of the HC$_3$N formation mechanism(s) \citep[\eg][]{tan17}.

Through a combination of laboratory, modeling and observational studies \citep[\eg][]{cla97,wil03,tea08}, it has been shown that Titan's nitriles (including HCN and HC$_3$N) are likely to become incorporated into more complex polymers and aerosol particles. As a result, the preferential removal of $^{15}$N from the atmosphere, through its incorporation into photochemical products and subsequent precipitation onto the surface, should be considered as an important $^{15}$N loss process, and thus a possible factor in the time-evolution of Titan's bulk atmospheric \ration\ ratio. The detection of a strong $^{15}$N-enrichment in a second molecule (after HCN) confirms the likely importance of this fractionation process.  The consequent increase in precipitation rate for atmospheric $^{15}$N (relative to $^{14}$N) means that the \ration\ ratio in N$_2$ may have been lower in the past \citep[\eg][]{kra16}, perhaps close to the value of 136 found in cometary NH$_3$ ice \citep{shi16}.

Our high-resolution HC$_3$N maps also reveal new details on Titan's atmospheric dynamics. By virtue of its large abundance at high altitude, its strong rotational emission spectrum and short lifetime ($\lesssim1$~yr; \citealt{wil04,kra09}), \hc\ is an excellent tracer of Titan's seasonally-variable atmospheric circulation. Over the 24 month period since the 2015 April ALMA observations of \hc\ \citep{lai17}, the ratio of S to N polar emission peak intensities has increased by almost a factor of two (from 3.1 to 5.7). This is explained by the combination of (1) rapid photolytic breakdown of \hc\ in the northern-hemisphere due to the increase in solar insolation approaching the 2017 solstice and (2) the transport of freshly-synthesized \hc\ from mid-latitudes towards the south pole by the strengthening winter polar circulation system \citep{tea12,lor15}. Details regarding these photolytic and transport mechanisms may be elucidated through future monitoring of the \hc\ distribution at high resolution.

Comparison of the \hc\ $v=0$ and $v_7=1$ line strengths reveals a high opacity in the $v=0$ line, approaching complete saturation at the winter pole. From a fitted 2D Gaussian, the S-polar peak is $0.21''\times0.38''$ in size (unresolved on the short axis), and the relatively low peak line brightness temperature of 78~K (compared to the $\approx150$-180~K gas temperature) indicates that saturated emission fills less than half the beam. This demonstrates the presence of one or more extremely compact regions of enhanced \hc\ column density at the winter pole, similar to the chemically-enriched gas recently observed at $<-80^{\circ}$ latitude by Cassini CIRS \citep{vin17,tea17}.

\section{Conclusions}

We have detected and mapped for the first time $^{13}$C and $^{15}$N isotopologues of \hc\ in Titan's atmosphere, revealing a high resolution snapshot of the global distributions for these trace gases. Similar to the main isotopologue, the \hct\ and \hcf\ show compact emission peaks in the vicinity of the south pole, consistent with a short photochemical lifetime and advective transport by Titan's seasonally-variable atmospheric circulation cell.

Our derived \ration\ ratio in \hc\ of $67\pm14$ represents a significant enrichment in $^{15}$N compared with the bulk (precursor) N$_2$ reservoir, and is the second such molecule for which this effect has been observed, after HCN. Good agreement between our \hc\ observations and the latest chemical modeling work demonstrates a reasonable understanding regarding the synthesis of this molecule in Titan's atmosphere, and confirms the importance of isotope-selective N$_2$ photochemistry.  Additional chemical/dynamical modeling is needed to investigate the full extent of Titan's $^{15}$N and $^{14}$N sources and sinks, to help further constrain the detailed time evolution of the bulk \ration\ ratio, that may provide new insights into the origin of Titan's nitrogen atmosphere. Such models may be tested through comparison of predicted gas abundance distributions with future ALMA observations at higher sensitivity.

\acknowledgements
This work was supported by the NSF under Grant No. AST-1616306. It makes use of ALMA data ADS/JAO.ALMA\#2016.A.00014.S. ALMA is a partnership of ESO, NSF (USA), NINS (Japan), NRC (Canada), NSC and ASIAA (Taiwan), in cooperation with the Republic of Chile. The Joint ALMA Observatory is operated by ESO, AUI/NRAO, and NAOJ. NRAO is a facility of the NSF operated under cooperative agreement by AUI. CAN and SBC were supported by the NASA HQ Science Innovation Fund. NAT was funded by the UK Science and Technology Facilities Council.

%%%%%%%%%%%%%%%%%%%%%%%%%%%%%%%%%%%%%%%%%%%%%%%%%%



%\bibliography{../article1/TitanRefs} % if your bibtex file is called example.bib

%%%%%%%%%%%%%%%%%%%%%%%%%%%%%%%%%%%%%%%%%%%%%%%%%%

%%%%%%%%%%%%%%%%% APPENDICES %%%%%%%%%%%%%%%%%%%%%

%%%%%%%%%%%%%%%%%%%%%%%%%%%%%%%%%%%%%%%%%%%%%%%%%%

\end{document}